\documentclass[showpacs, pra,twocolumn,preprintnumbers ,amsmath, amssymb, superscriptaddress, aps]{revtex4-2}
\usepackage{pgfplots}
\usepackage{tikz}
\pgfplotsset{compat=1.17}
\usepackage{color}
\usepackage{amsmath,amssymb}
\usepackage{pifont}% http://ctan.org/pkg/pifont
\usepackage{bbold}
\usepackage{float}

\usepackage[caption=false]{subfig}
\usepackage{makecell}
\usepackage{subfig}
\usepackage{graphicx} % Include figure files
\graphicspath{{Figures/}}
\usepackage{dcolumn}  % Align table columns on decimal point
\usepackage{bm}       % bold math
\usepackage{multirow} % Table functions
\usepackage{placeins}% For making floats not move around everywhere
\usepackage[colorlinks]{hyperref}
\hypersetup{colorlinks=true,linkcolor=red}
\usepackage{mathtools}
\usepackage{appendix}

\def \be{\begin{align}}
	\def \ee{\end{align}}
\def \bea{\begin{eqnarray}}
	\def \eea{\end{eqnarray}}

\begin{document}
	
\title{QLLVM: A Scalable Quantum-Classical Co-Compilation Framework based on LLVM}

\author{Yu Zhu}
\email{i_zhuyu@126.com}
\affiliation{The Laboratory for Advanced Computing and Intelligence Engineering, Zhengzhou 450001, China}

\author{Qiming Du}
\affiliation{The Laboratory for Advanced Computing and Intelligence Engineering, Zhengzhou 450001, China}

\author{Yuqiong Jin}
\affiliation{The Laboratory for Advanced Computing and Intelligence Engineering, Zhengzhou 450001, China}

\author{Woji He}
\affiliation{The Laboratory for Advanced Computing and Intelligence Engineering, Zhengzhou 450001, China}

\author{Hang Lian}
\affiliation{The Laboratory for Advanced Computing and Intelligence Engineering, Zhengzhou 450001, China}

\author{Xin Zhou}
\affiliation{The Laboratory for Advanced Computing and Intelligence Engineering, Zhengzhou 450001, China}

\author{Jianyu Zhang}
\affiliation{The Laboratory for Advanced Computing and Intelligence Engineering, Zhengzhou 450001, China}

\author{Yiyang Chen}
\affiliation{The Laboratory for Advanced Computing and Intelligence Engineering, Zhengzhou 450001, China}

\author{Jinchen Xu}
\email{atao728208@126.com}
\thanks{Corresponding author.}
\affiliation{The Laboratory for Advanced Computing and Intelligence Engineering, Zhengzhou 450001, China}

\author{Zheng Shan}
\email{shanzhengzz@163.com}
\thanks{Corresponding author.}
\affiliation{The Laboratory for Advanced Computing and Intelligence Engineering, Zhengzhou 450001, China}
			
\begin{abstract}				
To address the urgent need in the NISQ era for high-performance, scalable quantum compilers and to advance the integration of classical and quantum computing, we present QLLVM, an advanced Quantum-Classical co-compilation framework built on LLVM. To our knowledge, QLLVM delivers an end-to-end, LLVM-based compilation workflow that unifies the build of classical high-performance programs—including CUDA, MPI, and C++—together with quantum programs into a single executable. For quantum program compilation, QLLVM adopts a three-stage design: high-level optimizations are implemented in the MLIR Quantum dialect and then lowered to QIR—an LLVM IR–based representation—for low-level optimization and hardware mapping. Its extensible architecture and seamless interoperability with classical high-performance computing provide an efficient, flexible, industrial-grade compilation infrastructure for future quantum software development. Experimental results show that, on the MQTBench benchmark suite, QLLVM reduces circuit depth and gate counts compared with state-of-the-art compilers and demonstrates clear advantages in compiling hybrid classical–quantum programs.			
\end{abstract}
			
\maketitle

	%%%%%%%%%%%%%%%%%%%%%%%%%%%%%%%%%%%%%%%%%%%%%%%%%%%%%%%%%%%%
	\section{Introduction}
	%%%%%%%%%%%%%%%%%%%%%%%%%%%%%%%%%%%%%%%%%%%%%%%%%%%%%%%%%%%%%%%%%%
    	Quantum computing has emerged as a new computational paradigm with promising applications in drug discovery\cite{drug1,drug2,drug3,drug4,drug5,drug6,drug7}, materials design\cite{materials1,materials2,materials3,materials4,materials5,materials6,materials7}, cryptography\cite{cryptography1,cryptography2,cryptography3,cryptography4,cryptography5,cryptography6}, and artificial intelligence. Yet current devices remain limited by gate fidelities, qubit connectivity, and coherence times, keeping practical computation in the noisy intermediate-scale quantum (NISQ)\cite{file1} regime. As a consequence, achievable circuit depth and the number of effective logical qubits fall short of fault-tolerant requirements. Against this backdrop, high‑performance and scalable quantum compilers are a critical bridge between algorithmic theory and physical hardware. Such compilers must not only reduce gate counts and circuit depths at the gate level, but also couple tightly to classical control and numerical computation to meet the demands of increasingly prevalent classical–quantum hybrid workloads.

    Existing toolchains—such as Qiskit\cite{file2}, TKET\cite{file3}, and Cirq—have made substantial progress in circuit synthesis, gate optimization, and qubit mapping. However, the field grapples with two persistent challenges. First, the absence of a general, reusable compiler framework leads to redundant development efforts. Vendor-specific SDKs with tightly coupled compiler stacks result in fragmented intermediate representations (IRs), duplicated optimization passes, and high overhead for cross-platform portability. Second, integration with the classical high-performance computing (HPC) ecosystem is notably lacking. Weak compiler-level ties to established toolchains (e.g., LLVM, CUDA) and parallel programming models (e.g., OpenMP, MPI) hinder coordinated optimization and scheduling for hybrid quantum-classical applications.
    
    To address these gaps, particularly the integration with heterogeneous computing resources, unified programming models like CUDA-Q have been introduced. CUDA-Q provides a single framework for orchestrating CPU, GPU, and QPU execution, treating the quantum processor as a coprocessor tightly coupled with GPUs. This architecture delivers compelling performance for workloads with massive GPU-accelerated classical components. Nevertheless, its quantum–GPU co-processing paradigm is primarily tailored to GPU-centric workflows and does not natively address some CPU-centric HPC paradigms, such as MPI-based parallelism on large clusters. Furthermore, while CUDA-Q offers strong integration with GPU-based simulators, native support for a broad range of real quantum hardware topologies is still evolving.
    \begin{figure*}[ht]
	\centering
	\includegraphics[width=0.9\textwidth]{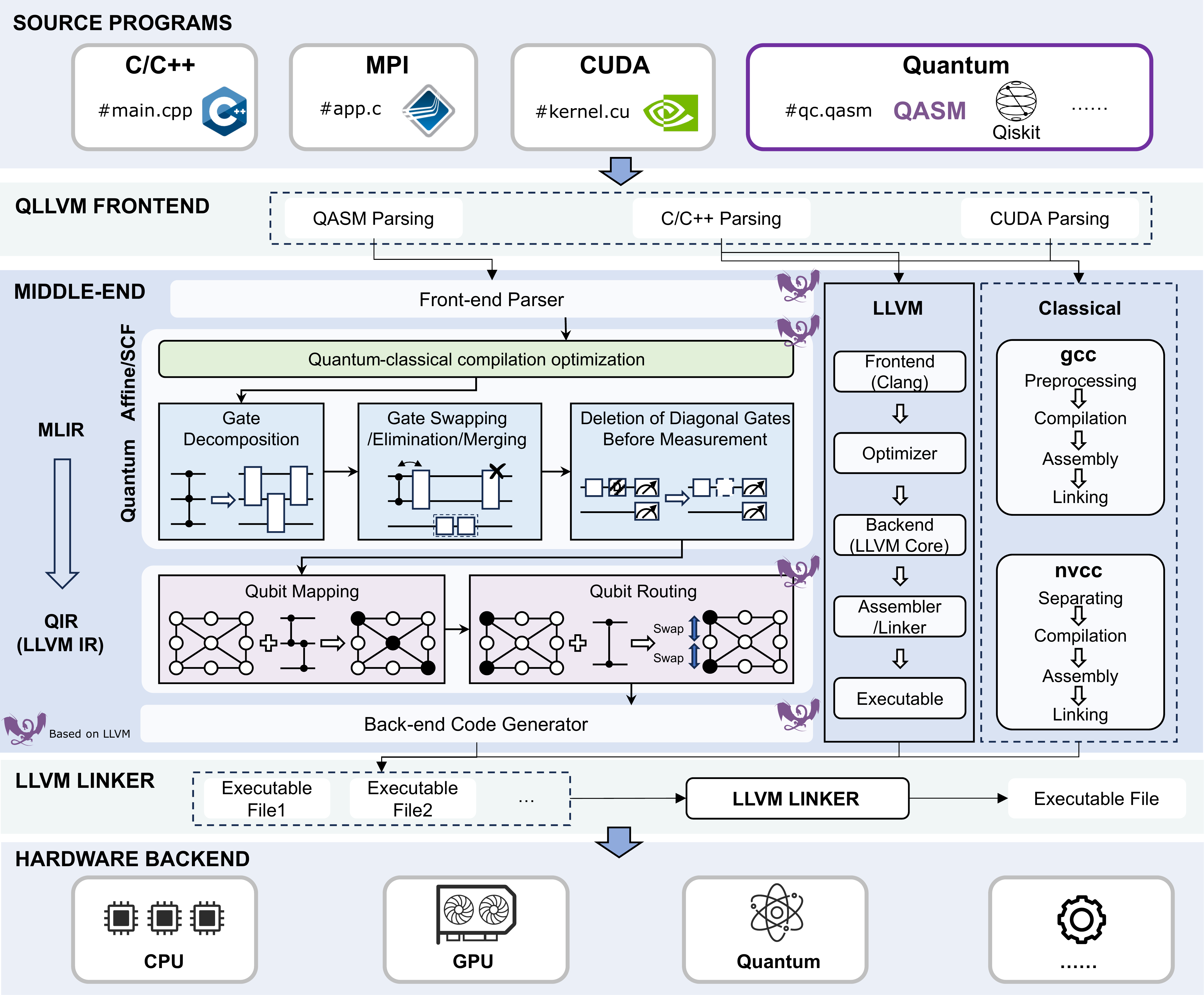}
	\caption{Collaborative Compilation Workflow of QLLVM for Hybrid Quantum-Classical Programs.}
	\label{fig:compilerframework}
	\end{figure*}
    
	In response, the community has pursued a unifying approach centered on intermediate representations. By constructing extensible IRs that span front ends and back ends—exemplified by QIR\cite{file10}—researchers aim to build unified quantum compilation frameworks that align classical and quantum semantics and catalyze ecosystem growth. Following this direction, frameworks such as ScaffCC and PennyLane have explored unified IRs and modular compilation pipelines. Nevertheless, LLVM/MLIR‑based frameworks still lack mature, hardware‑realistic solutions for physical qubit mapping and routing. For example, ScaffCC\cite{file7} and PennyLane\cite{file9} do not provide mapping and routing targeted at real devices; QIR‑based circuit manipulation and optimization can also suffer from low compilation efficiency. Moreover, deep compatibility between quantum program compilation and classical HPC compilation remains incomplete, limiting full‑stack optimization for classical–quantum workloads.

	To address these gaps, we present QLLVM, an extensible compiler framework centered on LLVM and MLIR. QLLVM unifies the compilation of classical HPC and quantum programs, while supporting multiple front ends and multiple quantum back ends. Distinct from existing LLVM/MLIR‑based quantum compilers, QLLVM performs hardware‑aware qubit mapping and routing at the LLVM IR stage (QIR), thereby enabling code generation for real devices within this design route.
    
    Leveraging the LLVM ecosystem, QLLVM integrates classical compiler passes, the CUDA programming model, and HPC runtimes to enable efficient compilation of hybrid workloads. Fig.~\ref{fig:compilerframework} illustrates the unified compilation pathway for classical–quantum programs. QLLVM accepts input programs including CUDA, C++, MPI, and quantum QASM programs. The QLLVM framework incorporates a hybrid program parsing module (frontend, analogous to LLVM Clang). Following hybrid program parsing, C/C++ and quantum programs are converted into a unified LLVM IR through the LLVM-based compilation flow. CUDA programs, after frontend parsing, are dispatched to the NVIDIA nvcc compiler for compilation, while MPI programs are processed via mpicc.
    
    The middle-end of QLLVM consists of optimization modules based on MLIR and LLVM IR, which serve as unified intermediate representations supporting both classical and quantum programs. For quantum programs, QLLVM implements a variety of quantum compilation and optimization techniques at the MLIR level, as well as qubit mapping and routing at the LLVM IR level. For classical programs, existing compilation optimizations from the LLVM infrastructure are applied at both the MLIR and LLVM IR levels. This unified intermediate representation enables tightly coupled classical and quantum computing. In the middle-end, CUDA programs are delegated to the CUDA nvcc compiler for compilation and processing.
    Upon completion of middle-end program optimizations, the backend performs code generation. Specifically, C/C++ and quantum programs are compiled via LLVM to generate corresponding executable binaries, while CUDA/MPI programs are compiled into target executables by nvcc and mpicc respectively. All components are finally assembled by the LLVM assembler into a unified executable. Upon execution of this unified binary, the corresponding program modules are dispatched to their respective target platforms for execution, and unified results are produced according to the invocation patterns among different modules within the hybrid program.
    
    \begin{figure*}[ht]
        \centering
        \includegraphics[width=0.8\textwidth]{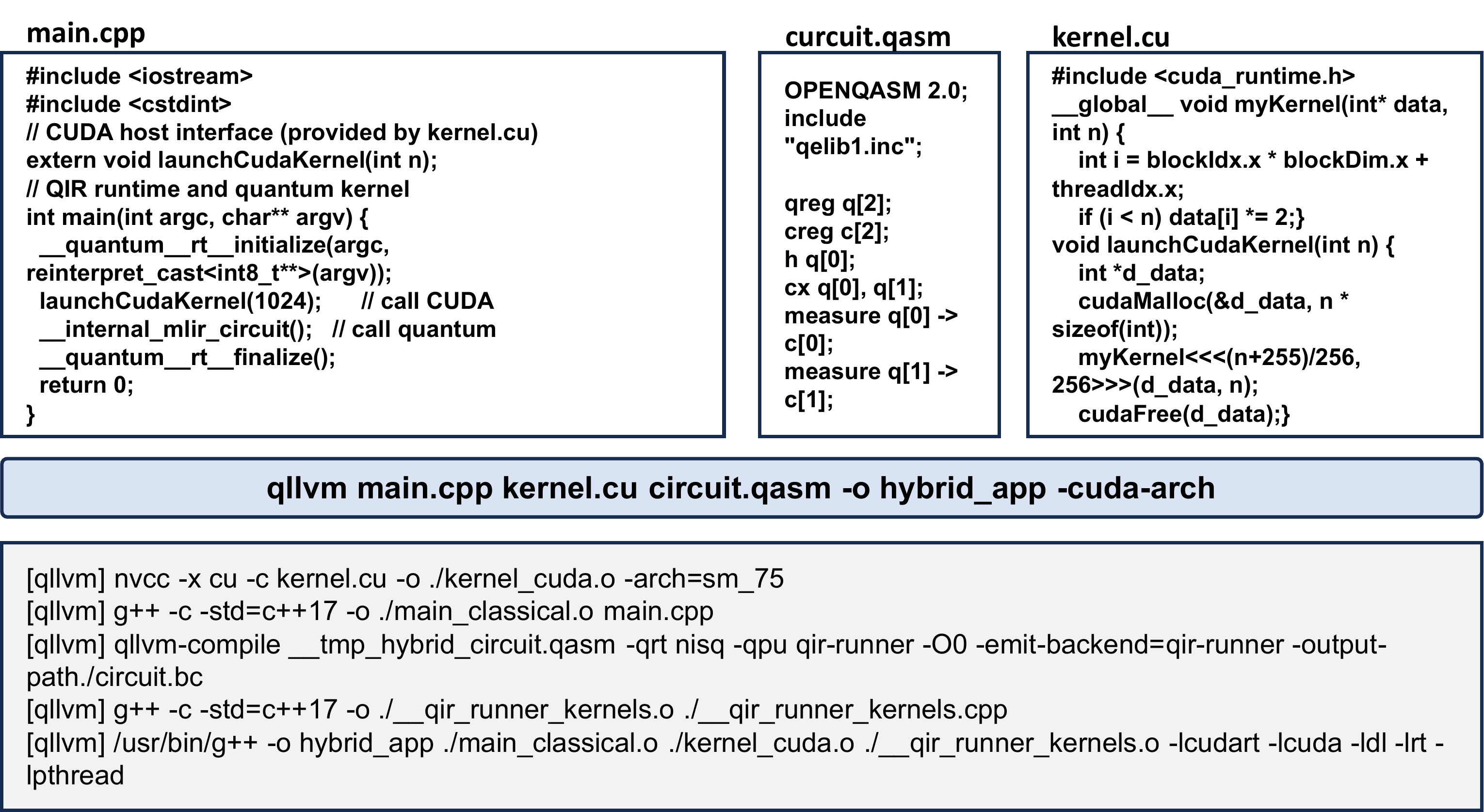}
        \caption{Example of Hybrid Compilation of Classical High-Performance Programs and Quantum Programs with QLLVM.}
        \label{fig:hybrid_code_example}
    \end{figure*}
    
    For quantum program compilation, the framework comprises front‑end language parsing, middle‑end optimization, and back‑end code generation. Quantum programs are first lowered to MLIR, which preserves high‑level constructs such as loop structures to support low‑overhead compilation of large‑scale circuits. After gate‑level optimizations at the MLIR level, QLLVM lowers MLIR to QIR (LLVM IR) and completes device‑specific qubit mapping and routing at this stage, finally emitting instruction code for different target back ends. This architecture supports multiple front‑end languages and back‑end platforms and affords flexible extensibility.
    
	We evaluate QLLVM along two axes: classical–quantum co‑compilation and quantum circuit optimization. In hybrid scenarios, through coordination with the LLVM toolchain and the QIR runtime, QLLVM achieves improved end‑to‑end compilation and execution performance. On public benchmarks such as MQTBench, compared with state‑of‑the‑art compilers, QLLVM further reduces circuit depth and gate counts. Taken together, these results provide a unified and reusable engineering foundation for high‑performance, scalable quantum compilation in the NISQ era, and a systems approach that deepens the integration of classical and quantum computing. The QLLVM project is openly available on GitHub at https://github.com/QCFlow/QLLVM.	
   
	%%%%%%%%%%%%%%%%%%%%%%%%%%%%%%%%%%%%%%%
	\section{Results}
	\label{sec:results}
	%%%%%%%%%%%%%%%%%%%%%%%%%%%%%%%%%%%%%%%
	
	This section evaluates QLLVM along two axes: (i) its ability to compile and execute classical–quantum hybrid programs in a unified toolchain, and (ii) its effectiveness in optimizing quantum circuits as compared to state-of-the-art compilers.
    
    Unless otherwise stated, all experiments are performed on an Intel Xeon Gold 6326 CPU server with 256 GB RAM and an NVIDIA A100 GPU (40 GB). QLLVM is built on LLVM 12.0.1 and MLIR 12.0.1. For comparison, we use Qiskit 1.2.4, Cirq 1.5.0, and PennyLane 0.42.3. All tools are configured to target the same universal gate set on an all-to-all connected logical topology, and are run with their respective “optimization level 3” or highest standard optimization setting. QLLVM is configured to enable all MLIR-level and QIR-level optimization passes, including single-qubit gate fusion and decomposition, but without any problem-specific manual tuning.
    
	\subsection{Collaborative compilation of Quantum-Classical hybrid programs}
	\label{sec:collab_comp}

	To assess QLLVM’s capability to serve as a unified compiler for hybrid workloads, we construct a representative benchmark that combines CPU-based control logic, a GPU-accelerated numerical kernel, and a quantum circuit. Figure~\ref{fig:hybrid_code_example} illustrates this collaborative compilation mechanism and provides a concrete example of the corresponding source code and compilation command. The hybrid application consists of three source modules: a C++ host program (main.cpp), a CUDA kernel (kernel.cu), and a quantum circuit specified in OpenQASM 2.0 (circuit.qasm). The C++ host first initializes the QIR runtime, calls the CUDA kernel to process a large array, and then invokes a quantum kernel compiled from circuit.qasm. The quantum kernel is executed on the qir-runner simulator backend.
    Using QLLVM, the entire hybrid application is compiled with a single command:
    % \verb|qllvm main.cpp kernel.cu circuit.qasm -o hybrid_app -cuda-arch <arch>|
    \begin{verbatim}
    qllvm main.cpp kernel.cu circuit.qasm
    -o hybrid_app
    -cuda-arch <arch>
    \end{verbatim}
    Internally, QLLVM orchestrates the following steps:
    \begin{itemize}
    \item It invokes the CUDA compiler (nvcc) to compile kernel.cu into a device-specific object file that embeds the GPU binary.
    \item It compiles the C++ host program main.cpp with an LLVM-compatible C++ compiler (g++) to produce an object file.
    \item It compiles circuit.qasm through the MLIR-based quantum frontend and midend to QIR bitcode (.bc) and generates a small runtime wrapper that loads and executes the QIR kernel on qir-runner.
    \item Finally, it links the three object files into a single executable hybrid\_app.
    \end{itemize}
	
	Execution of \texttt{hybrid\_app} subsequently dispatches the respective computational tasks to their designated backends (i.e., CPU, GPU, and quantum simulator) and aggregates the results into a cohesive final output.
	
	\subsection{Comparative analysis of compilation optimization}
	\label{sec:optimization_analysis}
	
\begin{figure*}[htbp]
		     \centering
		    \begin{tabular}{ccc}
                \includegraphics[width=0.314\textwidth]{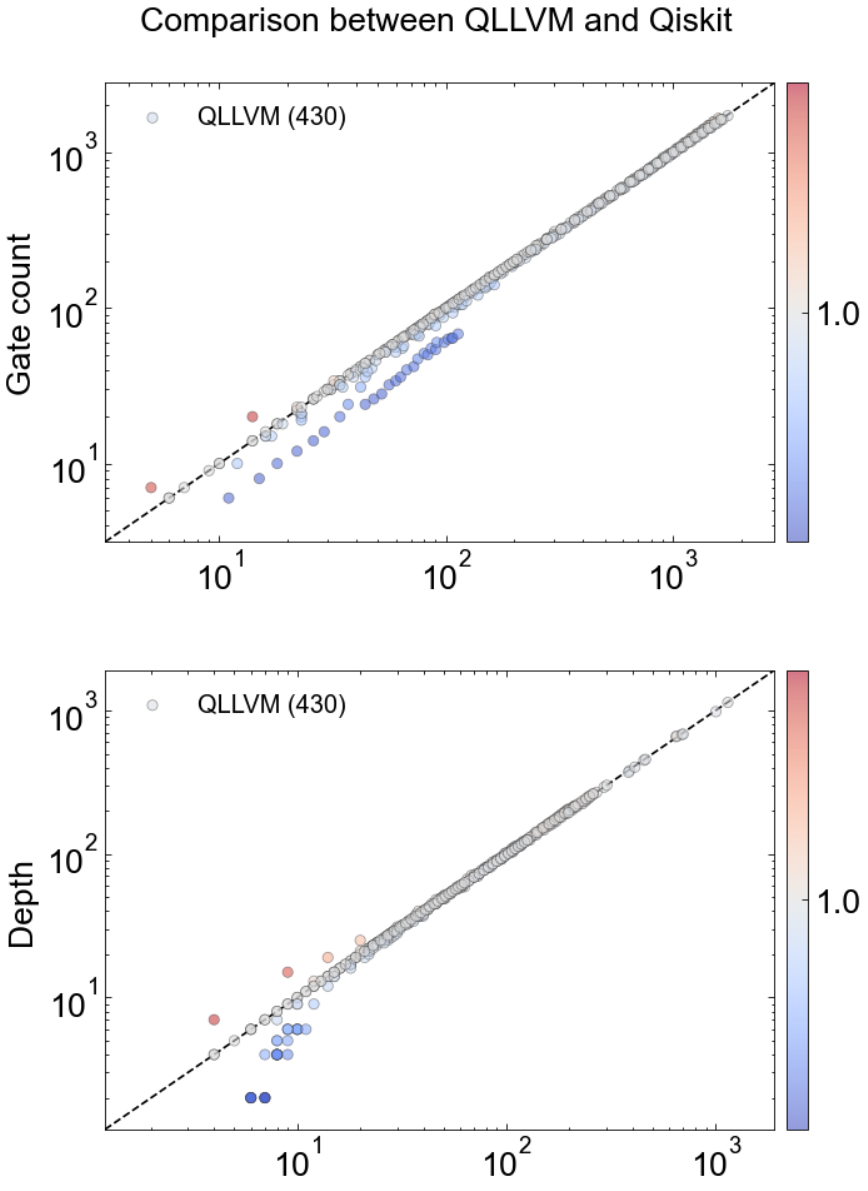} &
                \includegraphics[width=0.3\textwidth]{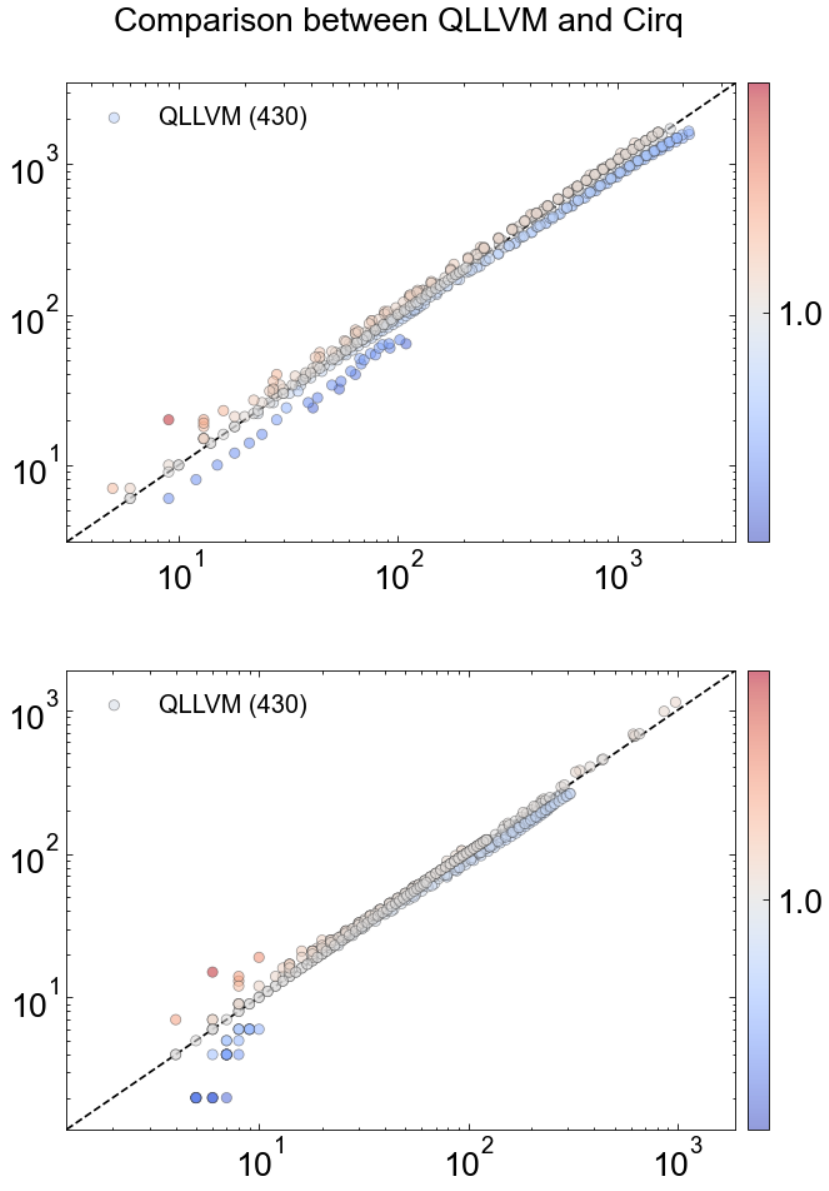} &
                \includegraphics[width=0.3\textwidth]{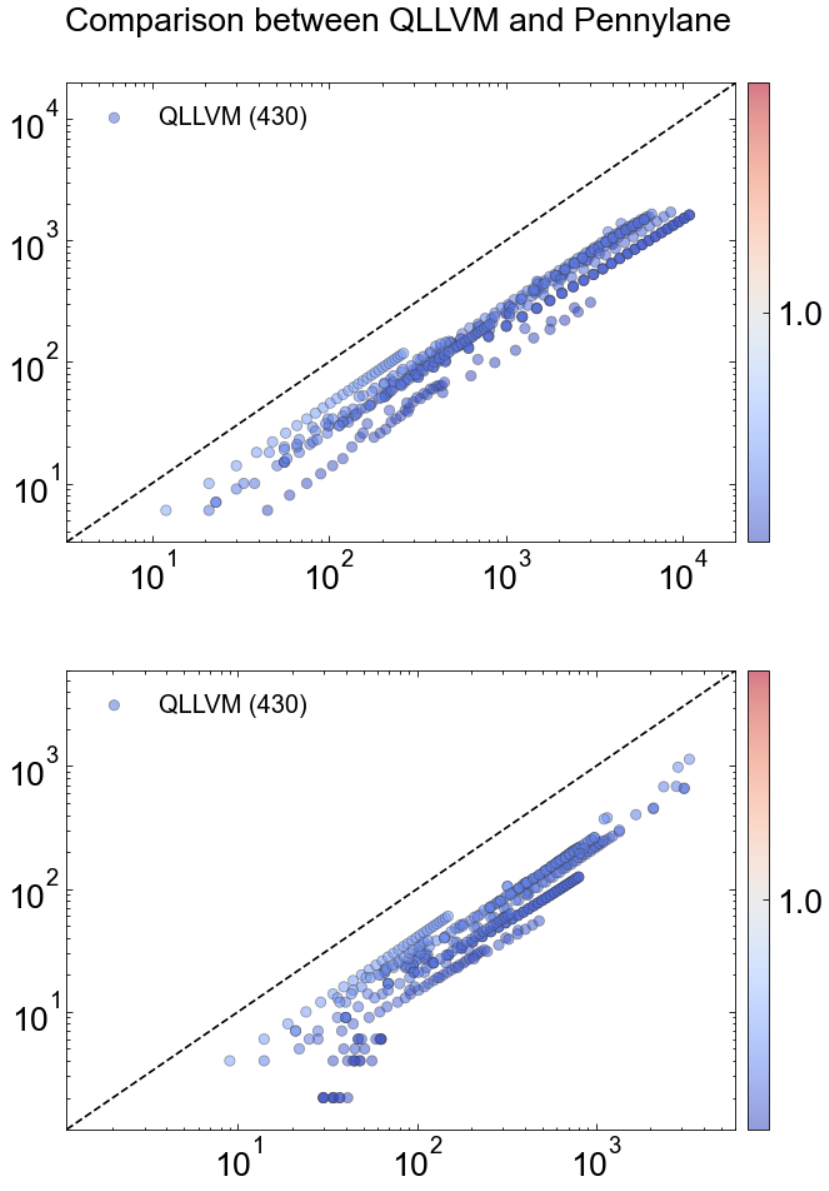} 
            \end{tabular}
		     \caption{Performance comparison between the QLLVM, Qiskit, Cirq, and PennyLane quantum compilers.}
		    \label{fig:performance-qllvm-vs-qiskit}
	\end{figure*}

    We benchmarked QLLVM's quantum optimization capabilities against leading compilers—Qiskit, Cirq, and PennyLane—using the comprehensive MQTBench suite \cite{file11}, which covers 25 algorithms with qubit counts from 3 to 30. Performance was measured by the final compiled circuit's gate count and depth, critical metrics for NISQ-era hardware.
    Fig.~\ref{fig:performance-qllvm-vs-qiskit} presents the comparison results of QLLVM with Qiskit, Cirq, and PennyLane in terms of performance indicators. Markers below the dashed line indicate tests where QLLVM performs better on a given metric relative to the specified compiler. By contrast, markers above highlight compiler performance that is better than that of QLLVM. 
    
    Specifically, QLLVM has similar optimization capabilities to Qiskit and Cirq. Compared to Qiskit, QLLVM achieves a 3.98\% reduction in gate count and a 3.56\% reduction in circuit depth on average across the benchmark suite, where averages are computed over all benchmark circuits with equal weight. Compared to Cirq, it reduces gate count by 1.19\% and circuit depth by 1.61\%. The improvements are even more pronounced when compared to PennyLane,with reductions of 74.96\% in gate count and 77.06\% in circuit depth, respectively.

	QLLVM's superior performance is primarily attributed to its advanced strategy for single-qubit gate optimization, which involves dynamic fusion and decomposition. During the compilation's optimization pass, QLLVM first fuses consecutive single-qubit gates into a single equivalent unitary matrix. Subsequently, it explores multiple decomposition pathways for this unitary matrix into standard forms, such as ZYZ, XYX, and ZXZ Euler decompositions. To ensure maximal optimization, the compiler dynamically selects the most compact representation based on an optimality criterion of minimizing the physical gate count, tailored to the native gate set of the target hardware. This method effectively reduces the number of single-qubit gates in the final circuit, leading to a direct reduction in circuit depth and a lower susceptibility to hardware noise.
    
	\section{Methods}
	%%%%%%%%%%%%%%%%%%%%%%%%%%%%%%%%%%%%%%%%%%%%%%%%
    This section describes the design and implementation of the Quantum-Classical co-compilation framework of QLLVM, its hybrid intermediate representation based on MLIR, and the qubit mapping and routing algorithms implemented at the LLVM IR level.
    
    Unless otherwise stated, QLLVM is implemented on top of LLVM 12.0.1 and MLIR 12.0.1, and adheres to the QIR 1.0 specification. All custom compiler passes described below are implemented as MLIR or LLVM passes and are available as part of the QLLVM source distribution.

	\subsection{Design and implementation of the quantum-classical co-compilation framework}
	\subsubsection{Unified driver and build semantics}
    QLLVM exposes a single driver, qllvm, that accepts a heterogeneous set of input source files and produces a single executable binary. The driver recognizes the following file types by extension:
    \begin{itemize}
    \item C/C++ sources (*.c, *.cc, *.cpp), compiled via an LLVM-compatible C/C++ compiler (clang++ or g++);
    \item CUDA sources (*.cu), compiled via nvcc to produce host object files that embed device binaries;
    \item MPI-enabled C/C++ sources, which are compiled via mpicc/mpicxx;
    \item Quantum sources (currently OpenQASM 2.0, and Qiskit circuits via an optional frontend), which are routed through the MLIR-based quantum frontend.
    \end{itemize}

    Given a mixed input such as \texttt{qllvm main.cpp kernel.cu circuit.qasm -o hybrid\_app -cuda-arch=sm\_75}, the driver executes the following steps:
    
    \begin{itemize}
    \item Classification. Each input file is classified by extension and compiler-specific flags. C/C++ sources are forwarded to clang++/g++; CUDA sources are compiled by nvcc with the specified -cuda-arch; quantum sources are handled by the QLLVM quantum frontend.
    \item Per-language compilation.
    \begin{itemize}
        \item C/C++ sources are compiled into ELF object files (*.o) using the standard LLVM compilation pipeline (frontend → optimizer → backend), with optimization level -O3 unless specified otherwise.
        \item CUDA sources are compiled by nvcc into host object files containing embedded device code; QLLVM passes through any user-specified nvcc flags unchanged.
        \item Quantum sources are lowered to MLIR, optimized by MLIR-level passes, and then translated to QIR bitcode (*.bc). For each quantum kernel, QLLVM generates a small C++ wrapper that exposes a callable function to the host and links it against the QIR runtime.
    \end{itemize}
    \item Linking. All object files from classical and quantum components, together with the QIR runtime and any required CUDA/MPI libraries, are linked by the LLVM-based linker (lld or the system linker) into a single executable. No special linker scripts are required; quantum kernels appear as normal externally visible functions to the linker.
    \end{itemize}
    
    Under this model, classical and quantum components are treated symmetrically: each language follows its native compilation pipeline to the level of object files or bitcode, and the final integration occurs at the link stage. This peer compilation model preserves compatibility with existing HPC build systems (e.g., CMake, Make) while enabling tight coupling with the quantum compilation path.
    
	\subsubsection{Integration with CUDA and MPI}
	For CUDA-enabled programs, QLLVM delegates device code generation entirely to nvcc. The driver invokes nvcc with the specified compute capability (e.g., -arch=sm\_75) and collects the resulting host-side object files. These objects are then linked with the rest of the application and with the CUDA runtime (libcudart, libcuda) in the final link step.
    
    MPI-based programs are supported by invoking mpicc/mpicxx for MPI sources. Each MPI rank may contain quantum kernels compiled by QLLVM; in this case, the QIR runtime is instantiated independently per rank. QLLVM does not alter the MPI execution model: program startup, rank creation, and message passing remain under the control of the MPI implementation. This design allows hybrid MPI+quantum programs to be compiled without modifying existing MPI code, while enabling each rank to invoke quantum circuits via the QIR runtime.
    
    QLLVM itself does not implement a distributed quantum runtime across MPI ranks; instead, it focuses on providing a compilation and linkage model in which MPI-based classical parallelism and quantum kernels can coexist in a single binary.

	\subsection{MLIR-based hybrid intermediate representation}
    \subsubsection{Quantum dialect design}
    At the core of QLLVM’s midend is a quantum dialect implemented on top of MLIR. The dialect provides a small set of operations and types that are sufficient to represent current quantum programs while remaining amenable to optimization and lowering to QIR.
    The quantum dialect introduces three primary types:
    \begin{itemize}
        \item Array: an ordered collection of opaque qubit references, corresponding to the QIR Array type;
        \item Qubit: an opaque handle to a single qubit, corresponding to the QIR Qubit type;
        \item Result: a measurement outcome, corresponding to QIR Result.
    \end{itemize}
 
    The dialect defines the following operations (see Table 1 for a summary):
    \begin{itemize}
        \item QRTInitOp / QRTFinalizeOp: initialize and finalize the QIR runtime, respectively;
        \item QallocOp / DeallocOp: allocate and deallocate quantum registers of type Array;
        \item QubitExtractOp: extract a Qubit handle from an Array at a given index (i64);
        \item InstOp: apply a quantum gate to one or more qubit operands, optionally parameterized by classical floating-point rotation angles, and optionally returning a Result.
    \end{itemize} 
    
    High-level frontends (such as an OpenQASM 2.0 parser) directly construct IR in this dialect. Loop constructs, conditional branches, and other control-flow structures are represented using MLIR’s standard SCF and Affine dialects, allowing classical and quantum control to coexist in a single IR.

    \subsubsection{Lowering from MLIR to QIR}
    Lowering to QIR is implemented as a dialect conversion from the quantum dialect to the MLIR LLVM dialect, followed by standard MLIR-to-LLVM code generation. Each quantum operation is mapped to a corresponding QIR runtime call:
    \begin{itemize}
      \item \texttt{QallocOp} is lowered to calls to \texttt{\_\_quantum\_\_rt\_\_qubit\_allocate\_array}.
      \item \texttt{DeallocOp} is lowered to calls to \texttt{\_\_quantum\_\_rt\_\_qubit\_release\_array}.
      \item \texttt{QubitExtractOp} is implemented as a call to
            \texttt{\_\_quantum\_\_rt\_\_array\_get\_element\_ptr} followed by a \texttt{bitcast} to \texttt{\%Qubit*}.
      \item \texttt{InstOp} is lowered to the corresponding \texttt{\_\_quantum\_\_qis\_\_*} function
            (e.g., \texttt{\_\_quantum\_\_qis\_\_h}, \texttt{\_\_quantum\_\_qis\_\_cx},
            \texttt{\_\_quantum\_\_qis\_\_rz}), determined by the gate name and its signature.
    \end{itemize}

    The conversion is implemented using MLIR’s ConversionPattern mechanism and a custom TypeConverter that maps quantum types (Array, Qubit, Result) to their QIR equivalents (opaque pointer types in the LLVM dialect). Unsupported gates are either decomposed into the target gate set at the MLIR level (e.g., breaking arbitrary single-qubit unitaries into rotations about X and Z) or, if no decomposition is provided, reported as compilation errors.
    
    After conversion, the IR is indistinguishable from hand-written QIR and can be processed by standard LLVM passes. Quantum-specific passes such as mapping and routing are implemented as LLVM IR passes that operate on functions with a designated “quantum kernel” attribute.

	\subsection{Qubit mapping and routing on LLVM IR}
    \subsubsection{Extracting circuit structure from QIR}
    Qubit mapping and routing in QLLVM operate on QIR functions after lowering from MLIR but before classical backend code generation. QLLVM identifies quantum kernels by scanning for functions that contain calls to QIS intrinsics (\texttt{\_\_quantum\_\_qis}). For each such function, a CircuitExtractor reconstructs the circuit structure from the low-level IR.
    The reconstruction proceeds in two stages:
    \begin{itemize}
        \item Logical-qubit identification. The extractor scans calls to QIR allocation and addressing functions (e.g., \_\_quantum\_\_rt\_\_qubit\_allocate\_array, \_\_quantum\_\_rt\_\_array\_get\_element\_ptr) and constructs a mapping from llvm::Value* instances representing qubit handles to logical qubit indices. This mapping is maintained per kernel function and assumes that qubit allocation and extraction follow the QIR specification.
        \item Gate sequence and dependency graph. The extractor then walks the basic blocks of the kernel function in program order, identifies calls to supported QIS gate functions (e.g., \_\_quantum\_\_qis\_\_h, \_\_quantum\_\_qis\_\_cx, \_\_quantum\_\_qis\_\_rz), and for each such call creates an internal Gate object. Each Gate records its kind (single-qubit or two-qubit), its operands (logical qubit indices), and a pointer to the originating CallInst. By maintaining, for each logical qubit, the last Gate that acted on it, the extractor constructs def–use chains and builds a directed acyclic graph (DAG) of gate dependencies. The set of gates with no unsatisfied predecessors forms the “front layer” used in the SABRE heuristic.
    \end{itemize}
    
    This process converts unstructured QIR into a gate list constrained by a DAG, decoupling subsequent mapping and routing algorithms from the specifics of LLVM IR.
    \subsubsection{SABRE-based mapping and routing}
    QLLVM adopts the SABRE algorithm to perform hardware-aware qubit mapping and routing. The algorithm is implemented as a function-level LLVM pass that takes as input a QIR kernel function and a target coupling graph, and produces a new function in which additional SWAP gates have been inserted to satisfy the connectivity constraints.

    The SABRE pass consists of three main components:
    \begin{itemize}
        \item Coupling graph construction. The target hardware topology is specified as an undirected graph G = (V, E), where vertices V represent physical qubits and edges E represent available two-qubit interactions. QLLVM stores G as an adjacency list and precomputes shortest-path distances between all pairs of physical qubits using repeated breadth-first search. These distances are used in the SWAP cost function.
        \item Initial layout selection. If no initial logical-to-physical qubit mapping is provided, QLLVM invokes the SabreLayout heuristic. Starting from a random permutation, the algorithm alternates between running the SabreSwap routing algorithm on the forward circuit and on the reversed circuit, updating the layout after each run. After a fixed number of iterations (three by default) or upon convergence, the layout that yields the minimum number of inserted SWAPs on the forward circuit is selected as the initial mapping.
        \item Front-layer SWAP heuristic routing. Given an initial mapping, QLLVM executes the circuit in topological order, maintaining the front layer F of ready gates. Two-qubit gates in F whose operands are mapped to adjacent physical qubits (i.e., connected by an edge in G) are executed immediately. For gates that violate adjacency, QLLVM generates candidate SWAP operations by considering pairs of physical neighbors adjacent to the mapped qubits.
    \end{itemize}
        
    Implemented at the LLVM IR level, SABRE inserts calls to the appropriate SWAP gate intrinsic (e.g., \_\_quantum\_\_qis\_\_swap) at the corresponding positions in the QIR function. Single-qubit gates and measurements are preserved, and any classical control flow present in the kernel (e.g., conditional branches based on measurement results) is currently assumed to be acyclic and is handled conservatively by restricting mapping to straight-line regions.

	\subsection{Discussion}
	Through deep integration with the LLVM/MLIR ecosystem, QLLVM provides a unified and extensible classical–quantum co-compilation stack. Its core contributions are compiler extensibility and seamless integration with classical high-performance computing (HPC) toolchains, delivering an efficient, flexible, and industrial-grade compilation infrastructure for future classical–quantum software development. The current implementation primarily targets the gate-model paradigm; just-in-time support for dynamically parameterized circuits in variational algorithms and hardware-specific optimizations for emerging architectures (e.g., neutral atoms and photonic platforms) remain limited. Compared with prior work, QLLVM advances the unification and extensibility of the compilation infrastructure; however, the breadth of algorithmic coverage and the depth of domain-specific optimization require further development. Future efforts will focus on supporting richer computational models, integrating machine-learning–driven adaptive optimizations, and deepening co-design with mainstream quantum control systems.

	\subsection{Acknowledgements}
	This work was supported by the National Key R\&D Program of China(No.2024YFB4504103).
	% \begin{thebibliography}{99}
		
	% \end{thebibliography}

\bibliographystyle{apsrev4-2}
\bibliography{Reference}	
	
\end{document}